\documentclass{article}
\usepackage[T1]{fontenc}
\usepackage[utf8]{inputenc}
\usepackage[lbd]{ismir} 
\usepackage{amsmath,cite,url}
\usepackage{graphicx}
\usepackage{color}
\usepackage{comment}

\usepackage[bookmarks=false,pdfsubject={\pdfsubject},hidelinks]{hyperref}

\title{A Traditional Approach to Symbolic Piano Continuation}

\multauthor
  {Christian Zhou-Zheng$^{1,2}$ \hspace{1cm} John Backsund$^2$ \hspace{1cm} Dun Li Chan$^3$ \hspace{1cm} Alex Coventry$^2$}
  {{\bf Avid Eslami$^4$ \hspace{0.8cm} Jyotin Goel$^5$ \hspace{0.8cm} Xingwen Han$^2$ \hspace{0.8cm} Danysh Soomro$^2$ \hspace{0.8cm} Galen Wei $^6$}\\
  $^1$ Metacreation Lab, Simon Fraser University \hspace{1cm} $^2$ EleutherAI\\
  $^3$ Sixth Form College, Universiti Sains Malaysia \hspace{1cm} $^4$ University of Toronto\\
  $^5$ Indian Institute of Technology Jodhpur \hspace{1cm} $^6$ Vanderbilt University\\
  {\tt\small christianzhouzheng@gmail.com}
  }

\def\authorname{C. Zhou-Zheng, J. Backsund, D. Li Chan, A. Coventry, A. Eslami, J. Goel, X. Han, D. Soomro, and G. Wei}

\begin{document}

\maketitle

\begin{abstract}
We present a traditional approach to symbolic piano music continuation for the MIREX 2025 Symbolic Music Generation challenge. While computational music generation has recently focused on developing large foundation models with sophisticated architectural modifications, we argue that simpler approaches remain more effective for constrained, single-instrument tasks. We thus return to a simple, unaugmented next-token-prediction objective on tokenized raw MIDI, aiming to outperform large foundation models by using better data and better fundamentals. We release model weights and code at \href{https://github.com/christianazinn/mirex2025}{https://github.com/christianazinn/mirex2025}.
\end{abstract}

\section{Introduction}\label{sec:introduction}

The generation of continuations of piano music has a long history in computational music generation, largely due to the propensity of piano music readily available in symbolic formats. Recent developments in sequence modeling have allowed continuation to be viewed as an autoregressive task, to be modeled with a suitable tokenization scheme and a powerful sequence model like the ubiquitous Transformer~\cite{vaswani}. A nonexhaustive list of prior work in this vein includes the Music Transformer~\cite{musictransformer_symbolic}, Museformer~\cite{museformer_symbolic}, FIGARO~\cite{figaro}, and MuseCoco~\cite{musecoco_symbolic}.

Most research in symbolic music modeling has so far focused on generalizing these techniques to---and improving performance on---long-sequence, multitrack, multi-instrument, and/or text- or attribute-controllable generative tasks. Typically, specialized techniques must be developed for these \emph{foundation models} to handle these harder tasks, such as fine- and coarse-grained attention for long sequences~\cite{museformer_symbolic}, and text feature extraction techniques~\cite{figaro} and attribute augmentation~\cite{musecoco_symbolic} for controllability.

However, when restricted to a simple, short-form, and single-instrument task, such as that posed by the MIREX 2025 Symbolic Music Generation challenge~\cite{mirex2025}, these sophisticated techniques may not be necessary, or may even be excessively complicated for the task. We conjecture that a return to the language-modeling approach taken by the Music Transformer~\cite{musictransformer_symbolic}, consisting of straightforward next-token prediction on tokenized raw musical data without any augmentations, will outperform foundation models when trained specifically for this task.

\section{System description}\label{sec:main}

As indicated previously, an accurate intuition for our approach may be obtained by taking the pretraining methodology of a standard large language model and replacing domain-specific components with their symbolic music counterparts (e.g. tokenized text with tokenized MIDI). We structure the remainder of this extended abstract as a technical report on our methodology, in the interest of reproducibility for future work in this vein.

\subsection{Objective}\label{sec:main:objective}

We adopt the objective of the MIREX 2025 Symbolic Music Generation challenge~\cite{mirex2025}, paraphrased as follows:

\begin{quote}
Given a 4-measure piano prompt, with an optional pickup measure, generate a musically coherent 12-measure continuation. Assume all music is in 4/4 time and quantized to a sixteenth-note resolution.
\end{quote}

The input and output are JSON objects, containing \verb|"prompt"| and \verb|"generation"| keys, which in turn contain lists of objects of the form
\begin{verbatim}
    {
      "start": 16,
      "pitch": 72,
      "duration": 6
    },
\end{verbatim}
where \verb|"start"| ranges from 0-79 for the prompt and 80-271 for the generation, and \verb|"pitch"| ranges from 0-127, corresponding to MIDI pitch numbers. We model the music in MIDI instead of JSON for ease of use with existing computational music tooling, and convert between the formats using scripts provided in the competition baseline.

\subsection{Data selection and preprocessing}

We used the recently released \href{https://huggingface.co/datasets/loubb/aria-midi/commit/ecc820daaf807b35f51673747e649735ae859cf5}{Aria-MIDI dataset}~\cite{bradshawaria} for training, under the premise that better data is always one of the most effective ways to improve performance. We used the \verb|pruned| split, discarded examples with an \verb|audio_score| of less than 0.9, and held out 10\% of the remainder for validation. The dataset comprises Type 0 MIDI files automatically transcribed from solo piano performances, which are rarely quantized, so we quantized all files to a sixteenth-note resolution ahead of training.

We used the MidiTok library~\cite{miditok2021} to implement a simplified REMI tokenization~\cite{pmt_remi}, resulting in a vocabulary size of 228. In particular, we disabled note velocity encodings (\texttt{use\_velocities=False}) and the splitting of tokens at bars or beats (\texttt{encode\_ids\_splits="no"}). These settings were chosen to efficiently model the simplified musical representation specified by the competition, as described in Section~\ref{sec:main:objective}.

Each training sample was generated by loading a MIDI file from the dataset with the symusic library~\cite{symusic}, encoding it with our REMI tokenizer, and selecting a random range of 16 consecutive bars that contained at least 100 tokens, which was passed to the model. Training samples typically ranged from 200 to 1200 tokens in length.

\subsection{Model architecture}

We adopted a decoder-only RWKV-7 backbone~\cite{rwkv7} as our sequence model, as it provides better data efficiency and easier training in a resource-constrained environment over the quadratic Transformer~\cite{vaswani}. We adopted the "deep and narrow" strategy suggested by Tay \emph{et al.}~\cite{tay2022scale} and Zhou-Zheng and Pasquier~\cite{zhouzheng2025personalizable}, resulting in a model with 12 layers, a hidden dimension of 384, and a feedforward dimension of 1536, for about 20 million total parameters. This small size allowed remarkably fast training and inference in a resource-constrained consumer environment.

\subsection{Training}

We trained for 50 epochs on a single RTX 4090 using the RWKV-LM library~\cite{peng_bo_2021_5196578}, using the Adam optimizer~\cite{kingma2014adam}, weight decay of 0.1, a batch size of 32, and a sequence length of 1024, as we found that the training sequences rarely exceeded this length. We used a cosine learning rate scheduler from 1e-4 to 1e-5 and optimized with respect to the standard cross-entropy loss. Each choice was informed by official recommendations and the first author's previous experience~\cite{zhouzheng2025personalizable}. Training took 46 hours to complete.

For testing, we took 8 samples on each of 7 test prompts at intervals of 4 epochs and qualitatively analyzed them. We found the model from epoch 32 to perform best, as later checkpoints showed a slight decline in quality and earlier checkpoints seemed undertrained.

\subsection{Inference}

We used the rwkv.cpp library~\cite{rwkv_cpp} to inference. The sampling parameters were tuned as follows: temperature=1.0, top-p=0.95, repetition penalty=1.0, top-k=40. We published an end-to-end system as a Docker image, which can be run with the following bash script with command \verb|bash script.sh in.json out/ n_sample|:

\begin{verbatim}
#!/bin/bash
IN_ABS=$(realpath "$1")
OUT_ABS=$(realpath "$2")
mkdir -p $OUT_ABS
USER="christianzhouzheng"
IMAGE="$USER/rwkv-mirex:latest"
docker pull "$IMAGE"
docker run --rm \
    -v "$IN_ABS:/app/input.json:ro" \
    -v "$OUT_ABS:/app/output" \
    "$IMAGE"  "/app/input.json" \
    "/app/output" "$3"
\end{verbatim}

\section{Results}\label{sec:res}

For the competition, a double-blind subjective listening test was performed, in which each of four anonymized models generated 8 samples for each of 8 prompts; from these 8 samples, each team cherry-picked one "best" sample for use in the listening test. The models compared were our 20M RWKV, another entry, and two baselines: the 780M \emph{Large} Anticipatory Transformer~\cite{thickstun2024anticipatory} and 1.2B \emph{xlarge} MuseCoco~\cite{musecoco_symbolic} foundation models.

Each survey participant was asked to rate each sample on a scale from 1 to 5 in each of the four following metrics: \textbf{coherency}, how well the continuation aligns with the style of the prompt; \textbf{structure}, how effectively it maintains a natural form and phrasing; \textbf{creativity}, the degree of novelty without losing musical sense; and \textbf{musicality}, the overall musical quality. Twenty participants with diverse musical backgrounds responded to the survey, of which 14 completed all eight pages with an average completion time of 32 minutes. The results are reported in Table~\ref{tab:model_comparison}.


\begin{table}[h!]
\centering
\begin{tabular}{lcc}
\hline
Model & Coherency & Structure \\
\hline
RWKV (ours) & $3.57 \pm 0.10^a$ & $3.58 \pm 0.10^a$ \\
PixelGen & $2.39 \pm 0.10^c$ & $2.37 \pm 0.09^c$ \\
Anticipatory \cite{thickstun2024anticipatory} & $\mathbf{3.70} \pm 0.10^a$ & $\mathbf{3.69} \pm 0.09^a$ \\
MuseCoco \cite{musecoco_symbolic} & $3.11 \pm 0.10^b$ & $3.07 \pm 0.09^b$ \\
\hline
\end{tabular}
\begin{tabular}{lcc}
\hline
Model & Creativity & Musicality \\
\hline
RWKV (ours) & $3.26 \pm 0.10^a$ & $\mathbf{3.50} \pm 0.10^a$ \\
PixelGen & $2.85 \pm 0.09^b$ & $2.48 \pm 0.09^c$ \\
Anticipatory \cite{thickstun2024anticipatory} & $\mathbf{3.30} \pm 0.10^a$ & $3.45 \pm 0.10^a$ \\
MuseCoco \cite{musecoco_symbolic} & $3.08 \pm 0.09^{ab}$ & $2.95 \pm 0.09^b$ \\
\hline
\end{tabular}
\caption{Model performance comparison. Each entry is reported as mean $\pm$ SEM$^s$, where SEM stands for the standard error of mean. Superscript $s$ is a letter: within each column, different letters indicate significant differences ($p < 0.05$) according to a Wilcoxon signed-rank test.}
\label{tab:model_comparison}
\end{table}

As visible in Table~\ref{tab:model_comparison}, our 20M RWKV-7 model clearly beat the PixelGen submission and the MuseCoco model, and performed on par with the strong Anticipatory Transformer baseline, which has 39 times more parameters (780M). The results reinforce our longstanding intuitions that simple, small, task-specific models can still perform on par or better than much larger foundation models at the specific task in question, and that the sophisticated techniques of foundation models may not be necessary for strong single-task performance.

\section{Conclusion}

We presented a simple, traditional approach to piano music continuation, inspired by recent advances in language modeling. By leveraging state-of-the-art resources and techniques, such as the RWKV-7 architecture~\cite{rwkv7} and Aria-MIDI dataset~\cite{bradshawaria}, we demonstrated that large, generalized foundation models can be matched or outperformed by a smaller, specialized model with a strong foundation. We conclude that there remains a niche for purpose-trained \emph{small models} in an era of progression defined by scaling parameter counts, especially in the field of music.

\section{Acknowledgments}\label{sec:ack}

We thank the MIREX 2025 organizers and the Symbolic Music Generation task captains for providing us with this opportunity. We also thank the members of the Metacreation Lab for their feedback and support.

This extended abstract was produced by a project under the inaugural \href{https://www.eleuther.ai/soar}{EleutherAI Summer of Open AI Research} (SOAR), led by project captain Christian Zhou-Zheng. We thank the anonymous Discord user \verb|genetyx8| for coordinating SOAR, and to Stella Biderman and EleutherAI for generously providing compute for other experiments.

\bibliography{main}

\begin{thebibliography}{10}
\providecommand{\url}[1]{#1}
\csname url@samestyle\endcsname
\providecommand{\newblock}{\relax}
\providecommand{\bibinfo}[2]{#2}
\providecommand{\BIBentrySTDinterwordspacing}{\spaceskip=0pt\relax}
\providecommand{\BIBentryALTinterwordstretchfactor}{4}
\providecommand{\BIBentryALTinterwordspacing}{\spaceskip=\fontdimen2\font plus
\BIBentryALTinterwordstretchfactor\fontdimen3\font minus \fontdimen4\font\relax}
\providecommand{\BIBforeignlanguage}[2]{{%
\expandafter\ifx\csname l@#1\endcsname\relax
\typeout{** WARNING: IEEEtran.bst: No hyphenation pattern has been}%
\typeout{** loaded for the language `#1'. Using the pattern for}%
\typeout{** the default language instead.}%
\else
\language=\csname l@#1\endcsname
\fi
#2}}
\providecommand{\BIBdecl}{\relax}
\BIBdecl

\bibitem{vaswani}
A.~Vaswani, N.~Shazeer, N.~Parmar, J.~Uszkoreit, L.~Jones, A.~N. Gomez, L.~Kaiser, and I.~Polosukhin, ``Attention is all you need,'' in \emph{Proceedings of the 31st International Conference on Neural Information Processing Systems}, Red Hook, NY, USA, 2017, p. 6000–6010.

\bibitem{musictransformer_symbolic}
C.-Z.~A. Huang, A.~Vaswani, J.~Uszkoreit, N.~M. Shazeer, I.~Simon, C.~Hawthorne, A.~M. Dai, M.~D. Hoffman, M.~Dinculescu, and D.~Eck, ``Music transformer: Generating music with long-term structure,'' in \emph{International Conference on Learning Representations}, 2018.

\bibitem{museformer_symbolic}
B.~Yu, P.~Lui, R.~Wang, W.~Hu, X.~Tan, W.~Ye, S.~Zhang, T.~Qin, and T.-Y. Liu, ``Museformer: transformer with fine- and coarse-grained attention for music generation,'' in \emph{Proceedings of the 36th International Conference on Neural Information Processing Systems}, Red Hook, NY, USA, 2022.

\bibitem{figaro}
D.~von R{\"u}tte, L.~Biggio, Y.~Kilcher, and T.~Hofmann, ``{FIGARO}: Controllable music generation using learned and expert features,'' in \emph{The Eleventh International Conference on Learning Representations}, 2023.

\bibitem{musecoco_symbolic}
P.~Lu, X.~Xu, C.~W. Kang, B.~Yu, C.~Xing, X.~Tan, and J.~Bian, ``Muse{C}oco: Generating symbolic music from text,'' \emph{arXiv preprint arXiv:2306.00110}, 2023.

\bibitem{mirex2025}
G.~Xia, J.~Jiang, A.~Maezawa, Z.~Wang, Y.~Zhang, R.~Yuan, and J.~S. Downie, ``{MIREX},'' \url{https://www.music-ir.org/mirex/wiki/MIREX_HOME}, Future MIREX Team, 2025, part of ISMIR Conference.

\bibitem{bradshawaria}
L.~Bradshaw and S.~Colton, ``Aria-{MIDI}: A dataset of piano {MIDI} files for symbolic music modeling,'' in \emph{International Conference on Learning Representations}, 2025.

\bibitem{miditok2021}
N.~Fradet, J.-P. Briot, F.~Chhel, A.~El~Fallah~Seghrouchni, and N.~Gutowski, ``{MidiTok}: A python package for {MIDI} file tokenization,'' in \emph{Extended Abstracts for the Late-Breaking Demo Session of the 22nd International Society for Music Information Retrieval Conference}, 2021.

\bibitem{pmt_remi}
Y.-S. Huang and Y.-H. Yang, ``Pop music transformer: Beat-based modeling and generation of expressive pop piano compositions,'' in \emph{Proceedings of the 28th ACM International Conference on Multimedia}, New York, NY, USA, 2020, p. 1180–1188.

\bibitem{symusic}
Y.~Liao and Z.~Luo, ``symusic: A swift and unified toolkit for symbolic music processing,'' in \emph{Extended Abstracts for the Late-Breaking Demo Session of the 25th International Society for Music Information Retrieval Conference}, 2024.

\bibitem{rwkv7}
B.~Peng, R.~Zhang, D.~Goldstein, E.~Alcaide, X.~Du, H.~Hou, J.~Lin, J.~Liu, J.~Lu, W.~Merrill, G.~Song, K.~Tan, S.~Utpala, N.~Wilce, J.~S. Wind, T.~Wu, D.~Wuttke, and C.~Zhou-Zheng, ``{RWKV}-7 "{G}oose" with expressive dynamic state evolution,'' \emph{arXiv preprint arXiv:2503.14456}, 2025.

\bibitem{tay2022scale}
Y.~Tay, M.~Dehghani, J.~Rao, W.~Fedus, S.~Abnar, H.~W. Chung, S.~Narang, D.~Yogatama, A.~Vaswani, and D.~Metzler, ``Scale efficiently: Insights from pretraining and finetuning transformers,'' in \emph{International Conference on Learning Representations}, 2022.

\bibitem{zhouzheng2025personalizable}
C.~Zhou-Zheng and P.~Pasquier, ``Personalizable long-context symbolic music infilling with {MIDI-RWKV},'' \emph{arXiv preprint arXiv:2506.13001}, 2025.

\bibitem{peng_bo_2021_5196578}
B.~Peng, ``{RWKV-LM},'' \url{https://github.com/BlinkDL/RWKV-LM}, 2025.

\bibitem{kingma2014adam}
D.~P. Kingma and J.~Ba, ``Adam: A method for stochastic optimization,'' \emph{arXiv preprint arXiv:1412.6980}, 2014.

\bibitem{rwkv_cpp}
RWKV, ``rwkv.cpp,'' \url{https://github.com/RWKV/rwkv.cpp}, 2025.

\bibitem{thickstun2024anticipatory}
J.~Thickstun, D.~Hall, C.~Donahue, and P.~Liang, ``Anticipatory music transformer,'' \emph{Transactions on Machine Learning Research}, 2024.

\end{thebibliography}

\end{document}